\documentclass[twocolumn,twoside,slac]{revtex4}
\usepackage{graphicx}
\usepackage{fancyhdr}
\begin{document}

\title{Geant4 hadronic physics status and validation for large HEP detectors}


\author{J.P. Wellisch}
\affiliation{CERN, Geneva, Switzerland}

\begin{abstract}
Optimal exploitation of hadronic final 
states played a key role in successes of all recent collider experiment in HEP, 
and the ability to use hadronic final states will continue to be one of the 
decisive issues during the analysis phase of the LHC experiments. 

Monte Carlo techniques facilitate the use of hadronic final
states, and have been developed for many years. We will give a brief 
overview of the 
physics underlying hadronic shower simulation, discussing the three 
basic types of modeling; data driven, parametrization driven, and theory 
driven modeling at the example of Geant4. We will confront these different 
types of modeling with the 
stringent requirements posed by the LHC experiments on hadronic shower 
simulation, and report on the current status of the validation effort
for large HEP applications. We will address robustness, and CPU 
and physics performance evaluations.
\end{abstract}

\maketitle

\thispagestyle{fancy}


\section{Model Overview}
% List of all models with very brief description.
% Which models are present or in preparation.
The number of model currently provided with or in development in the context
of {\sc Geant4} is growing continuously. We give an 
enumeration of the current status, including a brief description for each
model.

\subsection{Modeling Total Cross-sections}
The total cross-sections for inelastic scattering, capture of neutral particles,
induced fission and elastic scattering have been carried over from
GEANT3.21\cite{geant3}. The software design in {\sc Geant4} allows to 
overload this
default with specialized data-sets. Custom data sets are provided for proton
induced reactions\cite{protonInduced}, neutron induced
reactions\cite{neutronInduced}, pion reaction cross-sections\cite{bara}, and ion spallation reactions\cite{ionInduced},
as well as neutron interactions at energies below 20~MeV.

\subsection{Modeling Final States}
In modeling final states, three basic types of models are distinguished; Models
that are predominantly based on experimental or evaluated data, models that
are predominantly based on parameterizations and extrapolation of experimental
data under some theoretical assumptions, and models that are predominantly based
on theory.

\paragraph{Data driven models:}
When experimental or evaluated data are available with sufficient coverage, the data 
driven approach is considered to be the optimal way of modeling. 
Data driven modeling is used in the context of neutron transport, photon evaporation, 
internal conversion, radioactive decay, capture final states, absorption at rest, and isotope production.
We also use data driven modeling in the calculation of the inclusive 
scattering cross-sections for hadron nuclear scattering.
Limitations exist at high projectile energies, for particles with short 
life-times, and for strange baryons, as well as the $K^0$ system. 
Theory based approaches are 
 employed to extract missing cross-sections from the measured ones, or, 
 at high energies, to predict these cross-sections.

The main data driven models in {\sc Geant4} deal with neutron and proton induced
isotope production, and with the detailed transport of neutrons at low energies.
The codes for neutron interactions are 
generic sampling codes, based on the ENDF/B-VI data format, and evaluated neutron
data libraries such as ENDF/B-VI\cite{ENDF}, JENDL3.3\cite{JENDL}, and FENDL2.2\cite{fendl}. Note that any
combination of these can be used with the sampling codes.
The approach is limited by the available data to neutron kinetic energies up to 20~MeV, 
with extensions up to 30~MeV or 150~MeV for some isotopes.

The data driven isotope production models that run in parasitic mode to the 
transport codes are based on the MENDL\cite{MENDL} data libraries for proton and
neutron induced production. They complement the
transport evaluations in the sense that reaction cross-sections and
final state information from the transport codes define the interaction rate and particle 
fluxes, and the isotope production model is used only to predict activation.

The data driven approach is also used to simulate photon evaporation and internal 
conversion at moderate 
and low excitation energies, and for simulating radioactive decay. 
Both codes are based  on the ENSDF\cite{ensdf} data of 
nuclear levels, and transition, conversion, and emission probabilities. 
In the case of photon evaporation the data are supplemented by a simple theoretical 
model (giant dipole resonance) at high excitation energies.

Finally, data driven modeling is used in the simulation
 of the absorption of particles coming to a rest, mainly for
$\mu^-$, $\pi^-$, $K^-$, and $\bar{p}$, in order to describe the fast, direct 
part of the spectrum of secondaries, and in the low energy part of the modeling of elastic
scattering final states in scattering off Hydrogen. 
 
\paragraph{Parameterized models:}
Parameterizations and extrapolations of cross-sections and interactions are 
widely used in the full range of hadronic shower energies, and for all kinds 
of reactions. In {\sc Geant4}, models based on this paradigm are available for low 
and high particle energies respectively, and for stopping particles. They are
exclusively the result of re-writes of models available from GEANT3.21,
predominantly GEISHA\cite{geisha}. They include induced fission, capture, and elastic 
scattering, as well as inelastic final state production.

\paragraph{Theory based models:}
Theory based modeling is the basic approach in many models that are provided by or in
development for {\sc Geant4}. 
It includes a set of different theoretical approaches to describing hadronic interactions,
depending on the addressed energy range and CPU constraints. 

Parton string models for the simulation of high
energy final states ({$\rm E_{CMS}>O(5~GeV)$}) are provided and in further development.
Both diffractive string excitation, and dual parton model or quark gluon string
model are used. String decay is generally modeled using
well established fragmentation functions. The possibility to use 
quark molecular dynamic is currently in preparation.

In the energy regime
below 5~GeV center of mass energy, intra-nuclear transport models are provided. 
For cascade type models a re-write of
HETC\cite{HETC} as well as INUCL\cite{INUCL} is in 
provided, as well as an implementation
of a time-like cascade\cite{KinModel}. For quantum molecular dynamics models, 
an enhanced version of UrQMD\cite{URQMD}, as well as various variants of ablation/abrasion models are being written.
 
Note that the cascade models are based on average geometrical descriptions of
the nuclear medium, and take effects like Pauli-blocking, coherence length and formation
times into account in a effective manner. Scattering is done as in the
QMD, with the possibility to use identical scattering implementations. 
The QMD models 
calculate the interaction Hamiltonian from two- and
three-body interactions of all particles in the system, and solve the Newtonian
equations of motion with this time-dependent Hamiltonian numerically. Scattering is done
using smeared resonance cross-sections, taking Pauli's principle into account by 
investigating local phase-space. 
The approach promises to give all correlations in the final state correctly, and has no 
principle limitations in its applicability at low energies. It is very CPU expensive.

At energies below O(100~MeV) we provide the possibility to use exciton based 
pre-compound models to describe the energy and angular distributions of the fast particles. In this
area one model is released, and one more is in preparation.

The last phase of a nuclear interaction is nuclear evaporation. 
In order to model the behavior of excited, thermalised nuclei, variants of the classical 
Weisskopf-Ewing model are used. Specialized improvements such 
as Fermi's break-up model for light nuclei, 
and multi-fragmentation for very high excitation energies are employed. 
Fission, and photon
evaporation can be treated as competitive channels in the evaporation model.

As an alternative for, among others, intra-nuclear transport, the chiral invariant phase-space decay
model CHIPS is in development. It is a quark-level 3-dimensional event generator for 
fragmentation of excited hadronic systems into hadrons, and is
expected to find applicability in a wide range of hadron and 
lepton nuclear interactions, once fully explored.

A theoretical model for coherent elastic scattering was added recently, using 
the Glauber model and a two Gaussian form for the nuclear density. This
expression of the density allows to write the amplitudes in analytic form. Note
that this assumption works only since the nucleus absorbs hadrons very strongly at
small impact parameters, and the model describes nuclear boundaries 
well.

For lepton nuclear interactions, muon nuclear interactions are provided. Here
the leptonic vertex is calculated from the standard model, and the hadronic
vertex is simulated using a suitable set of models from the above described.
Neutrino nuclear interactions will be added in
due course.

\section{Sample data driven models}
% Description of selected models; Say one plot
As an example of a data driven model, we briefly describe the models for
neutron and proton
induced isotope production. 
These models are running in parasitic mode to the {\sc Geant4}\cite{geant4}
transport models, and can be used in conjunction with any set of models for
final state production and total cross-sections. 
They have been written to allow for detailed isotope production
studies, covering most of the spallation neutron and proton energy spectrum.
They are based on evaluated nucleon scattering data for kinetic energies
below 20~MeV, and a combination of evaluated data and 
extrapolations at energies up to 100~MeV. The upper limit of
applicability of the model is 100~MeV nucleons kinetic energy.

The evaluated data libraries that are the basis of the {\sc Geant4} neutron
transport and activation library G4NDL0.2 are Brond-2.1\cite{Brond},
CENDL2.2\cite{CENDL}, EFF-3\cite{EFF}, ENDF/B-VI.0\cite{ENDF}, ENDF/B-VI.1,
ENDF/B-VI.5, FENDL/E2.0\cite{fendl}, JEF2.2\cite{JEF}, JENDL-FF\cite{JENDL}, 
JENDL-3.1, JENDL-3.2, and MENDL-2\cite{MENDL}.

The G4NDL selection was guided in large parts by the FENDL2.0 selection.
Additions to and small modifications of this selection were possible due
to the structure of the {\sc shape Geant4} neutron transport code and the use of the
file system to maximize the flexibility of the data formats. The inclusion of the
MENDL data sets is fundamental for these models.

\begin{figure}
  \begin{center}
    \includegraphics[width=65mm]{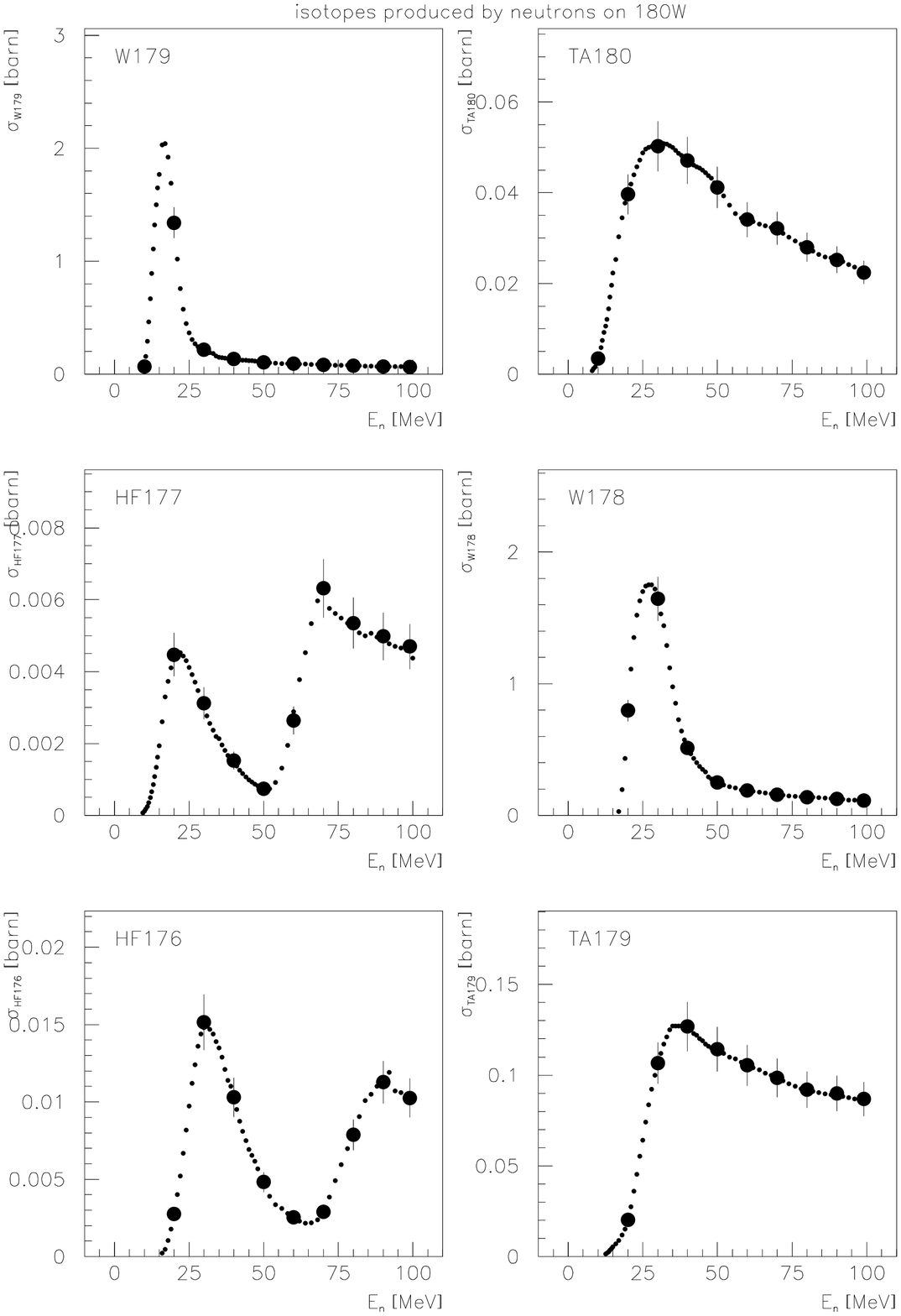}
    \caption{Isotope production cross-sections for neutron induced production of important
             isotopes as simulated using the isotope-production code in {\scshape Geant4}. Large 
	     points are simulation results, small points are evaluated data from the 
	     MENDL2 data library.}
    \label{fig:ex01}
  \end{center}
\end{figure}

Figure \ref{fig:ex01} shows an example of the simulated cross-section in comparison 
to evaluated data
from the MENDL collection, using $10^6$ events at each energy.
A systematic error of 15\% was added to the simulation results, to take the error  
in the extrapolation of the total cross-sections into account.
For a complete description and a more comparisons, see\cite{isotopes}.

\section{Sample parametrized models}
\begin{figure}
  \begin{center}
    \includegraphics[width=65mm, angle=270]{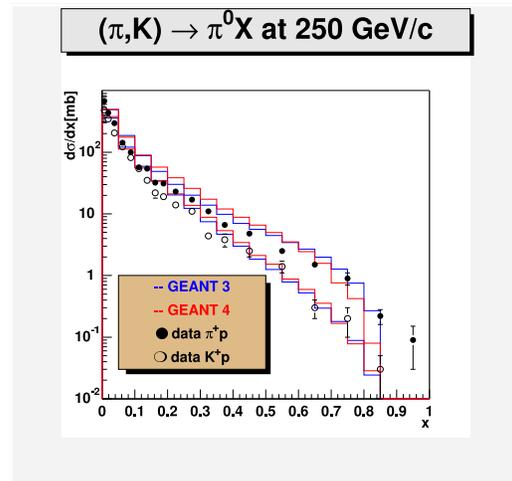}
    \caption{
Comparison of production cross-sections of neutral pions in kaon and pion 
induced reactions with measurement.
	    }
    \label{highenergy}   
  \end{center}
\end{figure}

% Description of selected models; one plot
Parameterization based models have been found to be very powerful in the case of
calorimeter simulation. Without giving a detailed description of these models,
we want to illustrate the predictive power for the case of the {\sc Geant4} high energy
models in Fig.~\ref{highenergy} for production of neutral pions in 
interactions of kaons and pions with Gold and Aluminum.
% Description of selected models; one plot
\section{Sample theory driven models}
\begin{figure}
  \begin{center}
    \includegraphics[width=65mm]{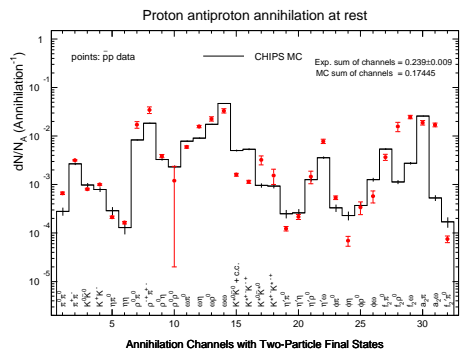}
    \caption{
Comparison of the branchings in two particle final states in proton
anti-proton annihilation with the predictions of CHIPS.
	    }
    \label{chips}   
  \end{center}
\end{figure}

% Description of selected models; one plot
% Use CHIPS here
Given that the chiral invariant phase-space decay model CHIPS is a rather new 
development and is developed only within {\sc Geant4}, we choose this as an 
example for a theory based model.
CHIPS is a quark-level 3-dimensional event generator for 
fragmentation of excited hadronic systems into hadrons. An important feature is
the universal thermodynamic approach to different types of excited hadronic
systems including nucleon excitations, hadron systems produced in
$e^{+}e^{-} $\ interactions, high energy nuclear excitations, etc.. Exclusive
event generation, which models hadron production conserving energy,
momentum, and charge, generally results in a good description of particle
multiplicities and spectra in multi-hadron fragmentation processes.
To illustrate the predictive possibilities of this ansatz, we show a comparison
between CHIPS predictions and measurement in the case of proton anti-proton
annihilation in Fig.~\ref{chips}.
For details of the model please see\cite{CHIPS1}, \cite{CHIPS2}, and \cite{CHIPS3}.
% Description of selected models; one plot
% Use CHIPS here
\section{Sample of a recent development}
The most recent developments in Geant4 hadronics concern the cascade codes. They have first been released for public use beginning 2003. They include a novel modeling Ansatz for cascade calculations, that we called Binary Cascade, which we find to have significant predictive power. It shall serve here as an example for a recent development.

Binary cascade introduces a new approach to cascade 
calculations, being based on a detailed 3-dimensional model of the nucleus, being based exclusively 
on binary scattering between reaction participants and nucleons within this nuclear model. In some 
sense this makes it a hybrid between a classical cascade code, and a quantum molecular dynamics 
model\cite{QMD}.

In binary cascading, like in QMD, each participating nucleon is described by a Gaussian wave-package.
$$\phi(x,q_i,p_i,t) = (2/(L\pi)^{3/4}exp(-2/L(x-q(t))^2+ip_i(t)x)$$
Here $x$, and $t$ are space and time coordinates, and $q_i$ and $p_i$ describe the particles' positions
in configuration and momentum space.

The total wave-function is assumed to be the direct product of the wave-functions of the participating
nucleons and hadrons, where participating means that they are either primary particles, or have been generated
or freed in the process of the cascade. Note that we do not take Slaters determinant into account in the
description. The wave function is not anti-symmetrized.

For such a wave-form, the equations of motion are identical in structure to the classical Hamiltonian
equations, and can be solved using the well known numerical integration methods of the cascade transport 
approach.

In binary cascade, unlike in QMD where it can be looked at as self-generating from the system configuration,
the Hamiltonian is calculated from optical potentials.

The imaginary part of the G-matrix acts like a scattering term. It is included in the model using discrete
scattering and particle decay in the cascade, with free 2-body cross-sections and a geometrical interpretation 
of the cross-section, and effective decay width for the strong resonances.

\begin{figure}
  \begin{center}
    \includegraphics[width=65mm]{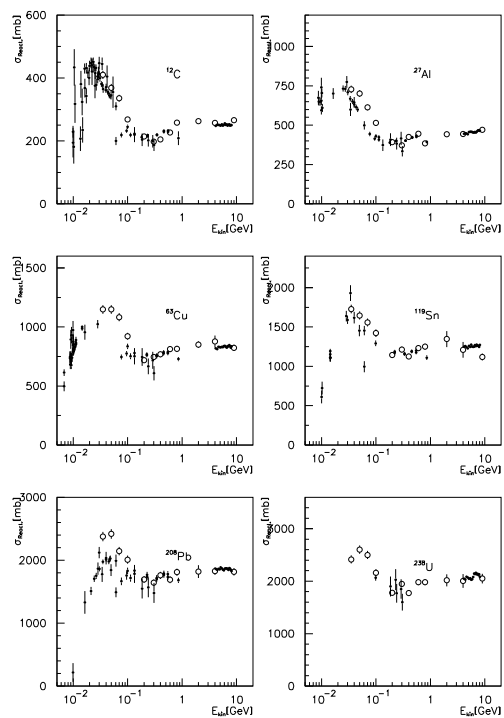}
    \caption{
Prediction for the total reaction cross-section in proton nuclear scattering for a set of nuclei as a function of the kinetic energy of the proton. Open circles are cascade predictions, and points are experimental data.
          }
    \label{bincross}   
  \end{center}
\end{figure}

\begin{figure}
  \begin{center}
    \includegraphics[width=65mm]{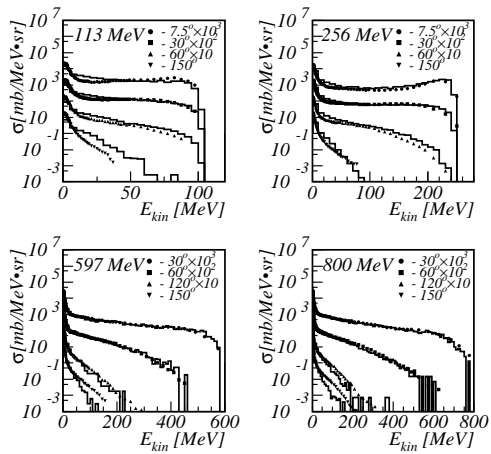}
    \caption{
Neutron spectra predicted by binary cascade for proton scattering off iron at a set of initial proton energies at various scattering angles. The histogram is Monte Carlo, and the points are experimental data.
          }
    \label{binfe}   
  \end{center}
\end{figure}

Two examples of the model's predictive power in proton nuclear scattering are given in Fig.~\ref{binfe} and Fig.~\ref{bincross}. Figure~\ref{bincross} shows the prediction for the total reaction cross-section in proton nuclear scattering for a set of nuclei as a function of the kinetic energy of the proton. The data are taken from reference\cite{cross}. Figure~\ref{binfe} shows the neutron spectra predicted by binary cascade  for proton scattering in iron at a set of initial proton energies at various scattering angles.
The data stem from the EXFOR database, and include data from references \cite{fe1}, \cite{fe2}, \cite{fe3}, and \cite{fe4}.
\section{Computational fundaments}
In this section  we give a brief impression of the usage of
Object Oriented frameworks for hadronic generators in {\sc Geant4}. 
We have put particular focus on the level of extendibility that can and
has been achieved by our Russia dolls 
approach to Object Oriented design, and implementation frameworks play a 
fundamental role in this.

A top-level, very abstracting implementation framework provides the basic
interface to the other {\sc Geant4} categories, and fulfills the most general
use-case for hadronic shower simulation, providing flexibility at the level of
selecting physics processes to be included in a simulation run.
It is refined for more specific use-cases by implementing a hierarchy of 
implementation frameworks, each level
implementing the common logic of particular use-cases, and refining the
granularity of delegation. Abstract classes are used as the delegation
mechanism\footnote{The same can be achieved with template specializations
with slightly improved CPU performance but at the cost of more complex designs
and less portability.}.
All framework functional and flexibility requirements were obtained through
use-case analysis. The lower level implementation frameworks address flexibility
in choice of cross-sections, models for final state production, and models for
isotope production (Level2), flexibility in the creation of theory driven models
from components like cascades, string-parton models, pre-equilibrium decay
models and evaporation phases (Level 3), flexibility of how to assemble cascade
or string parton models from components like scattering terms, string excitation, field
propagation, string fragmentation (Level 4), and flexibility and tailoring for fragmentation
functions in string decay(Level5). For details please see reference\cite{HPWCHEP2K}.

\section{Status of validation for large HEP detectors}
Much work has been invested by experimental groups to use and validate {\sc Geant4} hadronic physics for HEP detectors; in particular calorimetry, but also for tracker simulations. Very recently, these efforts were put onto a more solidly managed footing by the creation of the validation sub-project in the simulation project of the applications area of the LHC computing grid project. This effort is now being et up and lead by Fabiola Gianotti.

Prior validation efforts and usages of {\sc Geant4} hadronics in HEP we have become aware of are enumerated below:
\begin{itemize}
\item ATLAS tracker test-beam simulation; data with dedicated interaction trigger,
\item CMS tracker test-beam, cross-section validation,
\item ATLAS Tile Calorimeter test-beam simulation,
\item ATLAS Forward Calorimeter test-beam simulation,
\item ATLAS End-Cap Hadronic calorimeter test-beam simulation,
\item LHCb hadronic calorimeter test-beam simulation,
\item BTeV Crystal test-beam simulation,
\item CMS barrel combined test-beam simulation,
\item CsI/GLAST test-beam simulation,
\item H1 forward barrel test-beam simulation,
\item ATLAS combined end-cap simulation,
\item ALICE radiation protection benchmark simulation, and
\item CMS activation studies.
\end{itemize}
There is a strong possibility, that the above list is not exhaustive. It may not even be representative. Is meant solely to give an impression of the extent of usage of {\sc Geant4} hadronic physics in the HEP community.

In order to give an impression of the predictive power of {Geant4} hadronic physics, results of a repetition of one of the test-beam efforts using a simplified test-beam analysis are shown in Figs.\ref{etopi},\ref{res},\ref{shape}. They include data and simulation results from published, original sources, provided by the ATLAS end-cap community.
\begin{figure}
  \begin{center}
    \includegraphics[width=55mm,angle=-90]{epi.epsi}
    \caption{
Prediction of the e/pi ratio for the ATLAS end-cap calorimetry. Open circles are results of the simplified analysis, full points are results of the full analysis.
          }
    \label{etopi}   
  \end{center}
\end{figure}
\begin{figure}
  \begin{center}
    \includegraphics[width=65mm]{res.epsi}
    \caption{
Prediction of the energy resolution for the ATLAS end-cap calorimetry. Open circles are results of the simplified analysis, full points are results of the full analysis.
          }
    \label{res}   
  \end{center}
\end{figure}
\begin{figure}
  \begin{center}
    \includegraphics[width=65mm]{shape.epsi}
    \caption{
Prediction of the e/pi ratio for the ATLAS end-cap calorimetry. Open circles are results of the simplified analysis, full points are results of the full analysis.
          }
    \label{shape}   
  \end{center}
\end{figure}
\section{Conclusions}
% As usual, but without putting people in danger - the more collaboration
% the merrier it is, and CERN maintains the technical parts....
Taking the view of the LHC experiments, it has become evident that all modeling 
techniques - data driven, parameterization driven, and theory driven -
are need to satisfy all LHC needs in an optimal manner. Data 
driven modeling is known to provide
the best, if not only, approach to low energy neutron transport for radiation 
studies in
large detectors. Parametrization driven modeling has proven to allow for
tuning of the hadronic shower Monte Carlo for particle energies  
accessible to test-beam studies, is the most CPU performent possibility for calorimeter simulation, and is also widely used in this field. 
Theory driven modeling is the 
approach that promises safe extrapolation of results toward energies beyond the
test-beam region, and allows for maximal extendibility 
and customizability of the underlying physics.

The use of state of the art software technology is the key that allows for 
distributed development of the physics base of a hadronic shower simulation
tool-kit in the {\sc Geant4} context. 
It allows the work of many experts in the field to be combined in a
coherent manner, and offers the user the possibility to unify their knowledge 
in a single executable program in a manner that he deems optimal 
for his particular problem. 
This is a completely new situation. In a very short time it has lead 
to an unexpectedly wide range of modeling possibilities in {\sc Geant4}, 
and an unprecedented ease of flexibility of
usage of models and cross-sections. 

At the time of writing, {\sc Geant4} hadronic physics has become a very widely used program, with a predictive power in terms of physics that is as good or better than GEANT3 at its best.

\begin{acknowledgments}
The authors wish to thank the CERN IT and EP divisions for their support, and the HEP detector and experimental communities for their active collaboration in validating {\sc Geant4} hadronic physics. We especially thank the ATLAS end-cap community for being able to use their simulation results.
\end{acknowledgments}

\end{document}